\documentstyle[12pt]{article}
\newcommand{\beq}{\begin{equation}}
\newcommand{\eeq}{\end{equation}}
\newcommand{\beqn}{\begin{eqnarray}}
\newcommand{\eeqn}{\end{eqnarray}}
\newcommand{\stackM}{\stackrel{\scriptstyle >}{{ }_{\sim}}}
\newcommand{\stackm}{\stackrel{\scriptstyle <}{{ }_{\sim}}}  
\begin{document}

\thispagestyle{empty}
\def\pubnum{433}
\def\data{November, 1997}
\begin{flushright}
{\parbox{3.5cm}{
UAB-FT-433

November, 1997

hep-ph/9711472
}}
\end{flushright}
\vspace{3cm}
\begin{center}
\begin{large}
\begin{bf}
HEAVY CHARGED HIGGS BOSON DECAYING INTO TOP QUARK
IN THE MSSM\\
\end{bf}
\end{large}
\vspace{1cm}
J.A. COARASA, David GARCIA, Jaume GUASCH,\\
Ricardo A. JIM{\'E}NEZ, Joan SOL{\`A}\\
\vspace{0.25cm} 
Grup de F{\'\i}sica Te{\`o}rica\\ 
and\\ 
Institut de F{\'\i}sica d'Altes Energies\\ 
\vspace{0.25cm} 
Universitat Aut{\`o}noma de Barcelona\\
08193 Bellaterra (Barcelona), Catalonia, Spain\\
\end{center}
\vspace{0.3cm}
\hyphenation{super-symme-tric de-pen-ding}
\hyphenation{com-pe-ti-ti-ve}
\begin{center}
{\bf ABSTRACT}
\end{center}
\begin{quotation}
\noindent
Observing a heavy charged Higgs boson produced in
the near future at the Tevatron or at the LHC   would be
instant evidence of physics beyond the Standard Model.
Whether such a Higgs boson would be supersymmetric or not it could
only be decided after accurate prediction of its properties.
Here we compute the decay width of the dominant
decay of such a boson, namely $H^+\rightarrow t\,\bar{b}$, including
the leading electroweak corrections originating from large Yukawa couplings
within the MSSM. These electroweak effects turn out to be of comparable size
to the ${\cal O}(\alpha_s)$ QCD corrections in relevant portions of
the MSSM parameter space.
Our analysis incorporates the stringent low-energy constraints
imposed by radiative $B$-meson decays. 

\end{quotation}
  
\newpage

\baselineskip=6.5mm  

The Minimal
Supersymmetric extension of the Standard Model (MSSM) remains
nowadays as the only tenable Quantum Field Theory of the strong and the 
electroweak interactions beyond the SM that is able to keep pace with the
SM ability to (consistently) accommodate all known high precision
measurements\,\cite{WdeBoer}.
Moreover, the MSSM offers  
a starting point for a successful Grand Unified framework  where a
radiatively stable low-energy Higgs sector can survive.
All in all it is  well justified, we believe, to keep alive all
efforts on all fronts trying to discover a supersymmetric 
particle.  The next Tevatron
run, and of course also the advent of the LHC, should offer us a gold-plated
scenario for testing real, or at least virtual, manifestations of SUSY, if
this symmetry has anything to do at all with the origin of the electroweak
scale. 
A crucial part of the task aimed to understand the origin of this scale
is to unveil the nature of the spontaneous
symmetry-breaking mechanism and its likely connection
to a fundamental Higgs sector.

Thus, the less exotic -- and in this sense the most easily identifiable --
hint of SUSY physics would perhaps be the finding 
of a non-standard Higgs particle. It is well-known that
the MSSM predicts the existence of two charged Higgs 
pseudoscalar bosons, $H^{\pm}$,
one neutral CP-odd boson, $A^0$, and two neutral
CP-even states, $h^0$ and $H^0$ ($M_{h^0}< M_{H^0}$).
In the absence of direct sparticle detection, and because of the
similar phenomenological properties of the lightest neutral boson $h^0$
and the SM Higgs boson,  the {\it experimentum crucis} for the MSSM
could just be the discovery of a heavy charged
Higgs particle with accurate measurement
and prediction of its properties, namely at a level of quantum effects --
i.e. effects capable of revealing the details of the underlying supersymmetric
dynamics\,\footnote{For recent comprehensive reviews
of Higgs physics in the SM and MSSM, see e.g. Ref.\,\cite{Spira}.}. 
In connection to this possibility, we wish to show here that vestiges of 
virtual SUSY physics in the decay $H^+\rightarrow t\,\bar{b}$ can 
be large enough for even a hadron machine producing a heavy charged Higgs 
boson to be sensitive to them. 

As already emphasized in \cite{Ricard}, the $H^\pm\,t\,b$-vertex responsible for
the decay under consideration could be at the root of the Higgs production
mechanism itself. For, one expects that e.g. $H^+$ (similarly for $H^-$)
can be generously produced in hadron machines through $t\,\bar{b}$-fusion: 
$g\,g\rightarrow H^+\,\bar{t}\,b$ (Fig.\,1a)
as well as from charged Higgs bremsstrahlung off top and bottom
quarks\,\cite{Hstrahlung}:
$q\,\bar{q}\rightarrow H^+\,\bar{t}\,b$ (Fig.\,1b).
While the first mechanism is to be dominant at the LHC, the second one
could still give a chance to Tevatron, where Drell-Yan production
of $t\,\bar{t}$ and $b\,\bar{b}$ are the primary processes.  
In both cases one relies on the possibility of enhanced Yukawa couplings
of the charged Higgs boson with top and bottom quarks:
\begin{equation}
\lambda_t\equiv {h_t\over g}={m_t\over \sqrt{2}\,M_W\,\sin{\beta}}\;\;\;\;\;,
\;\;\;\;\; \lambda_b\equiv {h_b\over g}={m_b\over \sqrt{2}
\,M_W\,\cos{\beta}}\,.
\label{eq:Yukawas} 
\end{equation}
The process in Fig.\,1b is not necessarily
too suppressed against the ordinary two-body mode 
$q\,q^\prime\rightarrow W^{*}\rightarrow t\,\bar{b}$ as this
amplitude is purely electroweak, i.e. of ${\cal O}(\alpha_W)$, whereas the
former involves a three-body final state, but in compensation it is of order
${\cal O}(\alpha_s\,\lambda_b\sqrt\alpha_W)$; and so at large $\tan\beta$
(where $\lambda_b>1$) it may well afford a contribution 
of comparable size\,\cite{Hstrahlung}.

A preliminary supersymmetric treatment of $H^+\rightarrow t\,\bar{b}$  was put
forward in Ref.\,\cite{Ricard} (see also \cite{Bartl}), where
the ${\cal O}(\alpha_s)$ QCD effects were evaluated in
the MSSM\,\footnote{For the ordinary QCD and standard 
${\cal O}(\alpha_W\,m_t^2/M_W^2)$ corrections (in the $\lambda_b=0$
approximation) to that decay in a generic two-Higgs-doublet model
($2$HDM), see Ref.\,\cite{Oakes} and references therein.}.
However, to our knowledge, a thorough study within the MSSM including the
complete electroweak contributions from the Higgs boson
sector (with both $\lambda_t$ {\it and} $\lambda_b$ nonvanishing), together with
the host of sfermions and chargino-neutralinos, is not available
in the literature.
And this missing information can be essential for several
reasons. First, because the SUSY electroweak (SUSY-EW)
corrections could be enhanced
due to the intervention of supersymmetric top quark and bottom quark 
Yukawa couplings  of the type (\ref{eq:Yukawas}). 
Second, because for large gluino (and especially for large sbottom) masses
the SUSY-QCD corrections would no longer be 
that dominant\,\cite{Ricard}, and yet
potentially important
supersymmetric electroweak effects -- mainly sensitive to
stop and chargino exchanges -- could still
be alive\,\footnote{The SUSY-QCD and SUSY-EW 
corrections to the neutral Higgs boson decays into quarks
have been addressed in Ref.\,\cite{ToniRicard}.}.
However, these very same SUSY parameters are relevant to
the low-energy physics of the radiative $\bar{B}^0$-decays 
($b\rightarrow s\,\gamma$). Therefore, the severe constraints imposed by
this process cannot be ignored for the study of the charged Higgs decay,
and so we have taken them explicitly into account.
We have used -- and checked --the LO formula
(see the extensive literature\,\cite{Ng} for details):
\beq
BR (b\rightarrow s\,\gamma)\simeq BR (b\rightarrow c\,e\,\bar{\nu})\,
{(6\,\alpha_{\rm em}/\pi)\,\left(\eta^{16/23}\,A_{\gamma}+C\right)^2\over
I(m_c/m_b)\,\left[1-\frac{2}{3\pi}\,\alpha_s(m_b)f_{\rm QCD}(m_c/m_b)\right]}
\label{eq:bsg}\,,
\eeq
where
\beq
A_{\gamma}=A_{\rm SM}+A_{H^-}+A_{\chi^-\tilde{q}}
\label{eq:AS}
\eeq 
is the sum of the
SM, charged Higgs and chargino-squark
amplitudes, respectively. Although the NLO QCD corrections
to the SM ($W$-mediated) and charged Higgs mediated
amplitudes  are already
available (see e.g. Refs.\,\cite{Misiak,Ciucini}), still a $\sim 30\%$
uncertainty (similar to the  LO result in the SM) ought to be anticipated 
for the unknown MSSM contributions at the NLO.

A crucial issue concerning the SUSY-EW corrections is the
renormalization of $\tan\beta$. This parameter 
enters the lowest-order decay rate of $H^+\rightarrow t\,\bar{b}$
as follows: 
\beq \Gamma_0=\left({3\,G_F\,M_H^3\over
4\pi\sqrt{2}}\right)\, \lambda^{1/2} (1,x^2,y^2)
\left[(1-x^2-y^2)\,(x^2\cot^2\beta+y^2\tan^2\beta)
-4x^2\,y^2\right]\,,
\label{tree1}
\eeq
with $\lambda (1, x^2, y^2) = [1-(x+y)^2][1-(x-y)^2]$, and 
$x=m_t^2/M_H^2$, $y=m_b^2/M_H^2$, $M_H$ being the mass of $H^\pm$.
We shall follow the
procedure devised in Ref.\,\cite{CGGJS} where $\tan\beta$ is defined  by
means of the $\tau$-lepton decay of $H^\pm$:
\beq
\Gamma(H^{+}\rightarrow\tau^{+}\nu_{\tau})=
{\alpha m_{\tau}^2\,M_H\over 8 M_W^2 s_W^2}\,\tan^2\beta= 
{G_F m_{\tau}^2\,M_{H}\over 4\pi\sqrt{2}}\,\tan^2\beta\, 
(1-\Delta r^{MSSM})\,;
\label{eq:tbetainput}
\eeq
$\Delta r^{MSSM}$ is analyzed in \cite{Garcia}.
This definition generates a counterterm
\beq
{\delta\tan\beta\over \tan\beta}
=\frac{1}{2}\left(
\frac{\delta M_W^2}{M_W^2}-\frac{\delta g^2}{g^2}\right)
-\frac{1}{2}\delta Z_H
+\cot\beta\, \delta Z_{HW}+ 
\Delta_{\tau}\,.
\label{eq:deltabeta}
\eeq    
$\Delta_{\tau}$ above 
stands for the complete set of
MSSM one-loop effects on the $\tau$-lepton decay of $H^\pm$;
$\delta Z_{H}$ and $\delta Z_{HW}$ stand
respectively for the charged Higgs and mixed
$H-W$ wave-function renormalization factors; and the remaining 
counterterms $\delta g^2$ and $\delta M_W$ are the standard 
ones\,\cite{BSH}.
We would like to emphasize that the definition of $\tan\beta$
given above allows to renormalize the $H^{\pm}\,t\,b$-vertex in perhaps 
the most convenient way to deal with our main process
$H^+\rightarrow t\,\bar{b}$. Indeed,
from the practical point of view, we recall the excellent methods
for $\tau$-identification developed by the Tevatron
collaborations, which have recently been used by CDF to study the crossed
decay $t\rightarrow H^+\,b\, (\rightarrow\tau^+\,\nu_{\tau}\,b)$\,\cite{CDF}.
These techniques should prove very helpful to pin $\tan\beta$
down from experiment.

The general structure of the 
on-shell renormalized one-loop form factors in the MSSM is similar to
that in Refs.\cite{Ricard,CGGJS} and hence we shall refrain from 
exhibiting cumbersome analytical details\,\cite{Toni}.
Even though we shall explore the evolution of our results as a function
of the charged Higgs mass in the LHC range, 
for the numerical analysis we wish to single out the Tevatron accessible
window 
\beq
m_t\stackm M_H\stackm 300\,GeV\,.
\label{eq:interval}
\eeq
This window is especially
significant in that the CLEO measurements\,\cite{CLEO} of
$BR(b\rightarrow s\,\gamma)$ forbid most of this domain
within the context of a generic $2$HDM.
However, within the MSSM the mass interval
(\ref{eq:interval}) is perfectly consistent with eq.(\ref{eq:bsg}) 
provided that relatively light stop and charginos ($\stackm 200\,GeV$) 
occur\,\footnote{Although the inclusion of
the NLO effects on the charged Higgs corrected amplitude may considerably 
shift the range (\ref{eq:interval}) up to higher values 
of $M_H$\,\cite{Ciucini}, the NLO corrections on the SUSY amplitudes 
have {\it not} been computed, and so as in the LO case they might well
contribute to compensate the Higgs counterpart.}. We recall that for 
lighter chargino and stops ($\stackm 100\,GeV$) 
supersymmetric charged Higgs bosons may exist in the
kinematical window enabling the aforementioned top quark decay
$t\rightarrow H^+\,b$\,\cite{CGGJS,Guasch4}.

In Figs.\,2-5 we display in a nutshell our results
for a representative choice of parameters within
the present framework\,\footnote{See Ref.\,\cite{Toni}
for an exhaustive numerical analysis in the MSSM parameter space.}.
While in Figs.\,2-3 we have carefully determined a
region of the supersymmetric parameter space compatible with
the $b\rightarrow s\,\gamma$
measurements, in Figs.\,4-5 we exhibit the evolution of the quantum
corrections as a function of the most significant parameters.
To this end it will be useful to define the quantity 
\beq
\delta={\Gamma (H^+\rightarrow t\,\bar{b})-
\Gamma_0 (H^+\rightarrow t\,\bar{b})\over
\Gamma_0 (H^+\rightarrow t\,\bar{b})}\,,
\label{eq:deltag}
\end{equation}
which gives the correction with respect to the tree-level width (\ref{tree1}). 
The MSSM correction (\ref{eq:deltag}) includes the full QCD
yield (both from gluon and gluinos)
at ${\cal O}(\alpha_s)$ plus all the 
leading MSSM electroweak effects driven by the
Yukawa couplings (\ref{eq:Yukawas}).

Let us now elaborate a bit on the relevant region of the MSSM parameter space
that we have determined from the analysis of eq.(\ref{eq:bsg}).
This region (Cf. Figs.\,2-3) has been obtained in accordance with
the CLEO data\,\cite{CLEO} on radiative $\bar{B}^0$ decays at $2\,\sigma$.
Our set of independent MSSM inputs 
and remaining constraints is as in \cite{Ricard,CGGJS}; in particular,
we have imposed that non-SM contributions to the 
$\rho$-parameter be tempered by the relation
\beq
\delta\rho_{\rm new}\leq 0.003\,.
\label{eq:deltarho}
\eeq
Moreover, we have checked that the known
necessary conditions for the non-existence of
colour-breaking minima\,\cite{Frere} are fulfilled.
For definiteness, where $M_H$ has to be fixed, we have chosen
the value $M_H=250\,GeV$ within the range (\ref{eq:interval}), 
though we shall explicitly show the evolution of our results with $M_H$.
As for the dependence on the QCD renormalization scale $\mu_{QCD}$, 
following Ref.\,\cite{CLEO} we have entertained a variation of it
in the segment  $m_b/2\leq\mu_{QCD}\leq 2\,m_b$\, 
($m_b=5\,GeV$) and made allowance for
an additional $10\%$ theoretical
uncertainty. On the whole this amounts to a $\stackM 30\%$ 
indeterminacy in the MSSM prediction. Even so, the constraint
from  $b\rightarrow s\,\gamma$ in combination with the others does
project out a quite definite domain of the supersymmetric
parameter space. For example, 
in Fig.\,2a  we determine the allowed (shaded) region in the
$(\mu,A_t)$-plane for fixed values of the other parameters. 

The information from Fig.\,2a is indeed relevant since, as it is apparent 
in the plot, the trilinear coupling $A_t$ (a hot parameter modulating the
SUSY-EW corrections) becomes strongly correlated with 
the higgsino mixing parameter $\mu$, especially for low $\mu$.
The central vertical band around $\mu=0$ is excluded by our (conservative)
requirement that charginos should be heavier than $100\,GeV$.
 For $\mu<0$, we find $A_t>0$ in the permitted region
by $\bar{B}^0$ decays; conversely, for $\mu>0$, we find $A_t<0$. 
Similarly, in Fig.\,2b we plot the proper area in 
the $(\tan\beta,A_t)$-plane and we see that there exists 
a sizeable solution in the large $\tan\beta$ regime where to compute
$\Gamma (H^+\rightarrow t\,\bar{b})$.
There is of course a low $\tan\beta$ solution, too,
but in practice we shall only explore the large
$\tan\beta$ option. This is because the MSSM corrections (\ref{eq:deltag})
other than the ordinary QCD corrections are not significant
at low $\tan\beta$ (unless $\tan\beta<1$, which is not so appealing from 
the theoretical point of view) and thus in that circumstance
the potential SUSY nature of $H^\pm$ could not be disentangled from
the measurement of its top quark decay mode.
In the large $\tan\beta$ subdomain 
relevant to our Higgs decay, namely
\beq
20\stackm\tan\beta\stackm 50\,,
\label{eq:tansegment}
\eeq
the bottom quark Yukawa coupling, $\lambda_b$, 
is comparable to the top 
quark Yukawa coupling, $\lambda_t$\,\footnote{Theoretically, high values of
$\tan\beta$ as in eq.(\ref{eq:tansegment}) 
are well-motivated in the arena of widely different types of 
SUSY Yukawa coupling unification models\,\cite{GordosGiorgos}.}. 
In Fig.\,3 we describe the correlation with
the lightest sbottom and stop masses,
$m_{\tilde{b}_1}$ and  $m_{\tilde{t}_1}$.
Specifically, in Figs.\,3a and 3b we project the
$b\rightarrow s\,\gamma$
constraint onto the $(m_{\tilde{b}_1},A_t)$ and
$(m_{\tilde{t}_1},A_t)$ planes, respectively. From the first one it
is patent that there exists an 
essentially unlimited spectrum of heavy sbottom masses
compatible with any stop trilinear coupling in the range
$500\,GeV\stackm A_t\stackm 1\,TeV$ and 
without violating the $\delta\rho$ condition (\ref{eq:deltarho}) -- 
represented by the contour line hanging from above in Fig.\,3a.
This situation is different
from that in Fig.\,3b where there is a rather compact domain of proper 
$m_{\tilde{t}_1}$ values for each $A_t$. We emphasize that, contrary to 
the more commonly known result that holds at low $\tan\beta$, namely
that the lightest stop allowed by radiative $B$-meson decays
ought to be reachable at LEP$\,200$,
at high $\tan\beta$ the permissible values for $m_{\tilde{t}_1}$
are, instead, shifted away of the LEP$\,200$ possibilities. As a matter of fact,
the whole spectrum of sparticle masses that we use (including charginos)
is unreachable by LEP$\,200$. 

We are now ready to restrict our analysis of $H^+\rightarrow t\,\bar{b}$
within the appropriate domain pinpointed in Figs.\,2-3.  
We set out by looking at the branching ratio 
of $H^+\rightarrow\tau^+\,\nu_{\tau}$ (Cf. Fig.\,4).
Even though the partial width of this process does not
get renormalized (as it is used to define $\tan\beta$), its branching
ratio is seen to be very much sensitive to the MSSM corrections to
$\Gamma(H^+\rightarrow t\,\bar{b})$.
For large $\tan\beta$ as in eq.(\ref{eq:tansegment}), 
$BR(H^{+}\rightarrow\tau^{+}\nu_{\tau})$ may achieve rather high  
values ($10-50\%$) for Higgs masses in
the interval (\ref{eq:interval}), and it never
decreases below the $5-10\%$ level in the whole range.
Therefore, a handle for $\tan\beta$ measurement is always available
from the Higgs $\tau$-channel
and so also an opportunity for discovering quantum SUSY signatures on
$\Gamma (H^+\rightarrow t\,\bar{b})$.
As for the other $H^\pm$-decays, we note that the potentially important 
mode $H^{+}\rightarrow\tilde{t}_i\,\bar{\tilde{b}}_j$\,\cite{BartlMaj} does not
play any role in our case since (for reasons to be clear below) we 
are mainly led to consider bottom-squarks heavier than the 
charged Higgs.
Moreover, the $H^+\rightarrow W^+\,h^0$ decay which is sizeable enough at
low $\tan\beta$ becomes extremely
depleted at high $\tan\beta$\,\cite{Ricard}. Finally, the decays into 
charginos and neutralinos, $H^+\rightarrow \chi^+_i\,\chi_{\alpha}^0$, are not
$\tan\beta$-enhanced and remain negligible. 
Thus at the end of the day we do find an scenario
where $H^+\rightarrow t\,\bar{b}$ and $H^{+}\rightarrow\tau^{+}\,\nu_{\tau}$
can be deemed as the only relevant decay modes.

In order to assess the impact of the electroweak
effects, we demonstrate that a  typical set  of inputs
can be chosen such that the SUSY-QCD and 
SUSY-EW outputs are of comparable size.
In Figs.\,5a and 5b  we display
$\delta$, eq.(\ref{eq:deltag}), as a function respectively of $\mu<0$ 
and $\tan\beta$ for fixed values of the other parameters (within the 
$b\rightarrow s\,\gamma$ allowed region). Remarkably, in spite of the fact  
that all sparticle masses are beyond the scope of LEP$\,200$ the 
corrections are fairly large. 
We have individually plot the SUSY-EW, SUSY-QCD, standard QCD
and total MSSM effects. 
The Higgs-Goldstone boson corrections
(which we have computed in the Feynman gauge)
are isolated only in Fig.\,5b just to make clear that
they add up non-trivially to a very
tiny value in the whole range (\ref{eq:tansegment}), and 
only in the small corner $\tan\beta<1$  they can be of some significance.

In Figs.\,5c-5d we render the various corrections (\ref{eq:deltag})
as a function of the relevant squark masses.
For $m_{\tilde{b}_1}\stackm 200\,GeV$ we observe 
(Cf. Fig.\,5c) that the SUSY-EW contribution is non-negligible 
($\delta_{SUSY-EW}\simeq +20\%$) but 
the SUSY-QCD loops induced by squarks and gluinos are by far the
leading SUSY effects ($\delta_{SUSY-QCD}> 50\%$) -- the standard QCD
correction staying invariable over $-20\%$ and the standard EW correction
(not shown) being negligible. 
In contrast, for larger and larger $m_{\tilde{b}_1}>300\,GeV$, say
$m_{\tilde{b}_1}=400$ or $500\,GeV$, and fixed 
stop mass at a moderate value $m_{\tilde{t}_1}=150\,GeV$, the
SUSY-EW output is longly sustained whereas the SUSY-QCD one 
steadily goes down.
However, the total SUSY pay-off adds up to about
$+40\%$ and the net MSSM yield still
reaches a level around $+20\%$, i.e. of equal
value but opposite in sign to the conventional QCD result. This
would certainly entail a qualitatively distinct quantum signature.

We stress that the main parameter to decouple 
the SUSY-QCD correction is the lightest sbottom mass, rather than the
the gluino mass\,\cite{Ricard}.
For this reason,
since we wished to probe the regions of parameter space where
these electroweak effects are important, the direct 
SUSY decay $H^{+}\rightarrow\tilde{t}_i\,\bar{\tilde{b}}_j$ mentioned above is
blocked up kinematically and plays no role in our analysis.
On the other hand, the SUSY-EW output is basically controlled 
by the lightest stop mass, as it is plain in Fig.\,5d, where we
vary it in a range past the LEP$\,200$ threshold. 

We have also checked that in the alternative $\mu>0$, $A_t<0$ scenario 
(also admissible according to Fig.\,2a), the SUSY-QCD correction is negative
but it is largely cancelled by the SUSY-EW part, which stays positive, so
that the total $\delta_{MSSM}$ is negative and
larger (in absolute value) than the standard QCD correction. 
Finally, coming back to Fig.\,4 we remark that
if we take the  standard QCD-corrected branching ratio
(central curve in that figure)
as a fiducial quantity, rather than the corresponding tree-level result,
then $BR(H^+\rightarrow\tau^+\,\nu_{\tau})$ undergoes an effective 
MSSM correction of order\, $\pm (40-50)\%$. 
The sign of this effect is given by the
sign of $\mu$. In practice, $BR(H^+\rightarrow\tau^+\,\nu_{\tau})$ should
be directly measurable from the cross-section for 
$\tau$-production\,\cite{CDF}. 

To summarize, supersymmetric quantum effects on
the decay width of $H^+\rightarrow t\,\bar{b}$ could be sizeable enough to
seriously compete with the ordinary QCD corrections. 
Furthermore, our computation shows that
these effects are compatible with CLEO data from
low-energy $B$-meson phenomenology.
The present study completes preliminary supersymmetric
treatments where only the SUSY-QCD corrections
were calculated\,\cite{Ricard,Bartl} 
within the $(b\rightarrow s\,\gamma)$-unconstrained MSSM
parameter space.
Here we have evaluated for the first time the leading SUSY-EW effects 
and combined them with the SUSY-QCD ones both within the domain of 
compatibility with $b\rightarrow s\,\gamma$.
As a result, we confirm that also in the constrained case
the SUSY-QCD effects are generally very important\,\cite{Ricard}.
However, we have exemplified an scenario with sparticle masses above
the LEP$\,200$ discovery range where the SUSY electroweak corrections
triggered by large Yukawa couplings can be comparable to the
SUSY-QCD effects. In this context the total SUSY correction
remains fairly large --around $+(30-50)\%$-- with a $\sim 50\%$ 
component from electroweak supersymmetric origin.
This situation occurs for i) large $\tan\beta$ 
($>20$), ii) huge sbottom masses ($> 300\, GeV$) 
and iii) relatively light stop and charginos ($100-200\,GeV$).
If the charged Higgs mass lies in the 
intermediate window (\ref{eq:interval}),
a chance is still left for Tevatron to produce a charged Higgs
heavier than the top quark by means of
``charged Higgsstrahlung'' off top and bottom quarks.
Should, however, a heavier $H^\pm$ exist outside the window (\ref{eq:interval}),
the LHC could continue the searching task mainly from gluon-gluon fusion 
where again $H^\pm$ is produced in association with the top quark.
The upshot is that the whole range of charged
Higgs masses up to about $1\,TeV$ could be probed and, within the
present renormalization framework, its potential supersymmetric nature be 
unravelled through a measurement of $\Gamma (H^+\rightarrow t\,\bar{b})$ 
with a modest precision of $\sim 20\%$. Alternatively, one could look
for indirect SUSY quantum effects on
the branching ratio of $H^+\rightarrow \tau^+\,\nu_{\tau}$ 
by measuring this observable to within
a similar degree of precision.

{\bf Acknowledgements}:

\noindent
One of us (J.S.) thanks M. Carena, C. Wagner and G. Zoupanos for useful
discussions on  high $\tan\beta$ SUSY GUT's.
This work has been partially supported by CICYT 
under project No. AEN93-0474. The work of D.G. and J.G. has been financed by 
grants of the Comissionat per a Universitats i Recerca, Generalitat de 
Catalunya.


\vspace{0.75cm}
\begin{center}
\begin{Large}
{\bf Figure Captions}
\end{Large}
\end{center}
\begin{itemize}
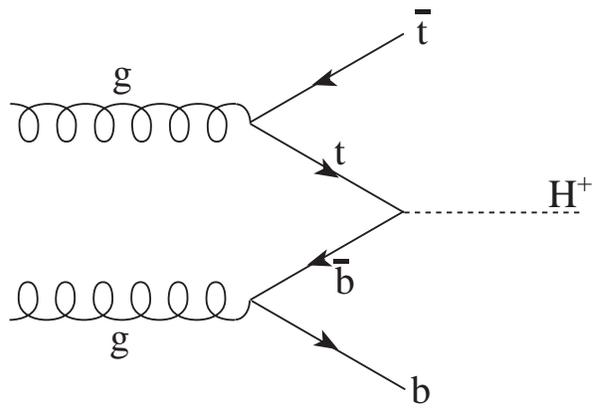
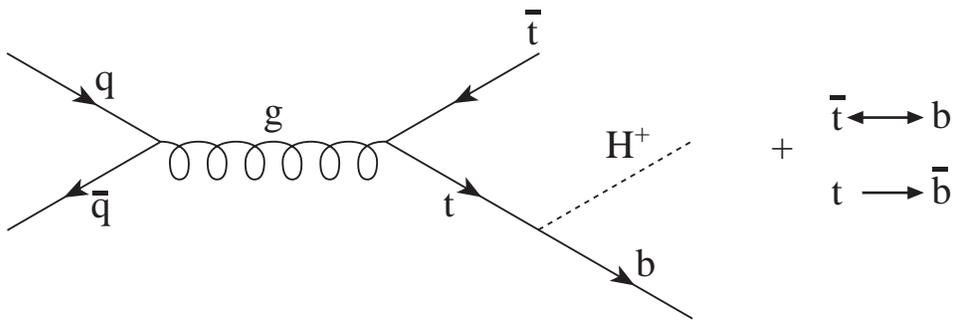
\item{\bf Fig.1} Typical charged-Higgs production mechanisms at hadron
colliders:
{\bf (a)} $H^+$ production through $t\,\bar{b}$-fusion; and
{\bf (b)} through charged Higgs bremsstrahlung off top and bottom
quarks.

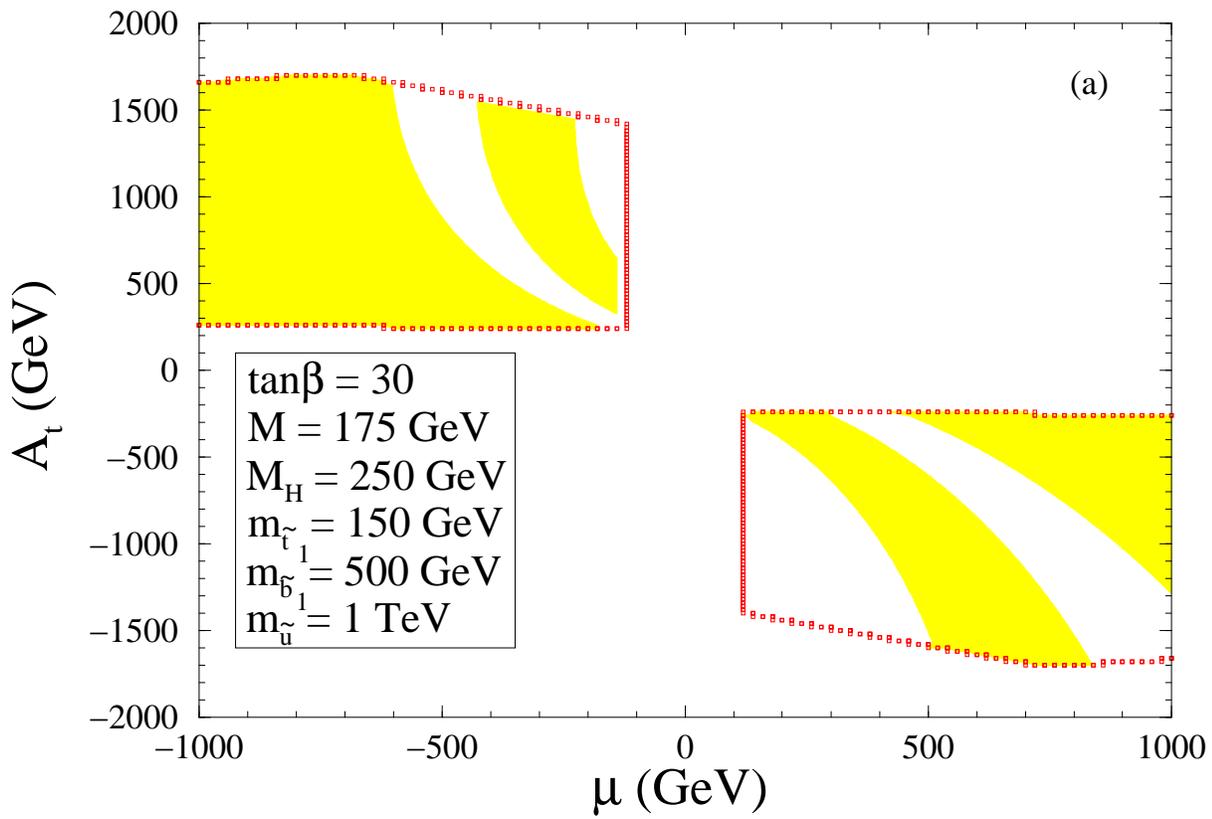
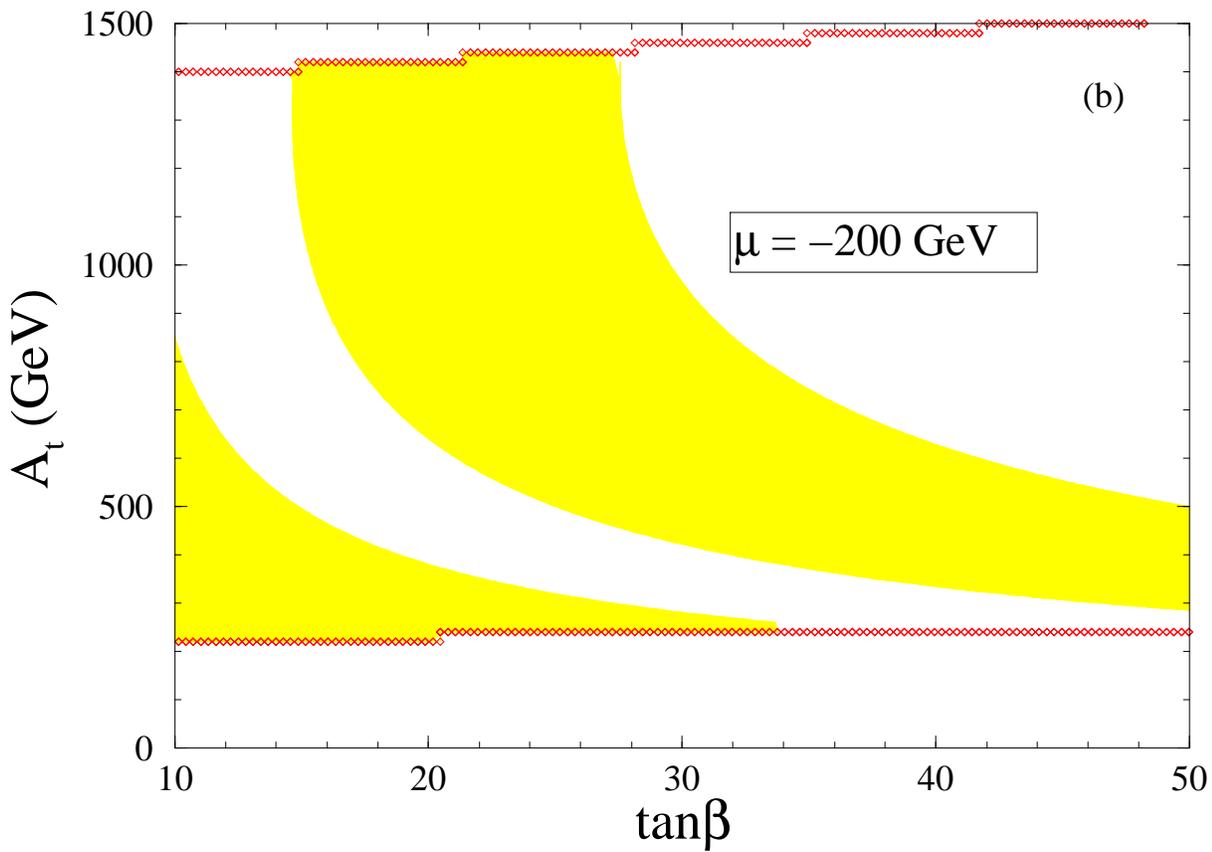
\item{\bf Fig.2} Domains of the MSSM parameter space allowed by
$b\rightarrow s\,\gamma$ at $2\,\sigma$ level and the 
theoretical constraints explained in the text, for given values
of the other parameters. {\bf (a)} Permitted region in the  $(\mu,A_t)$-plane;
{\bf (b)} In the $(\tan\beta, A_t)$ plane. 
The proper domains are the shaded ones. 

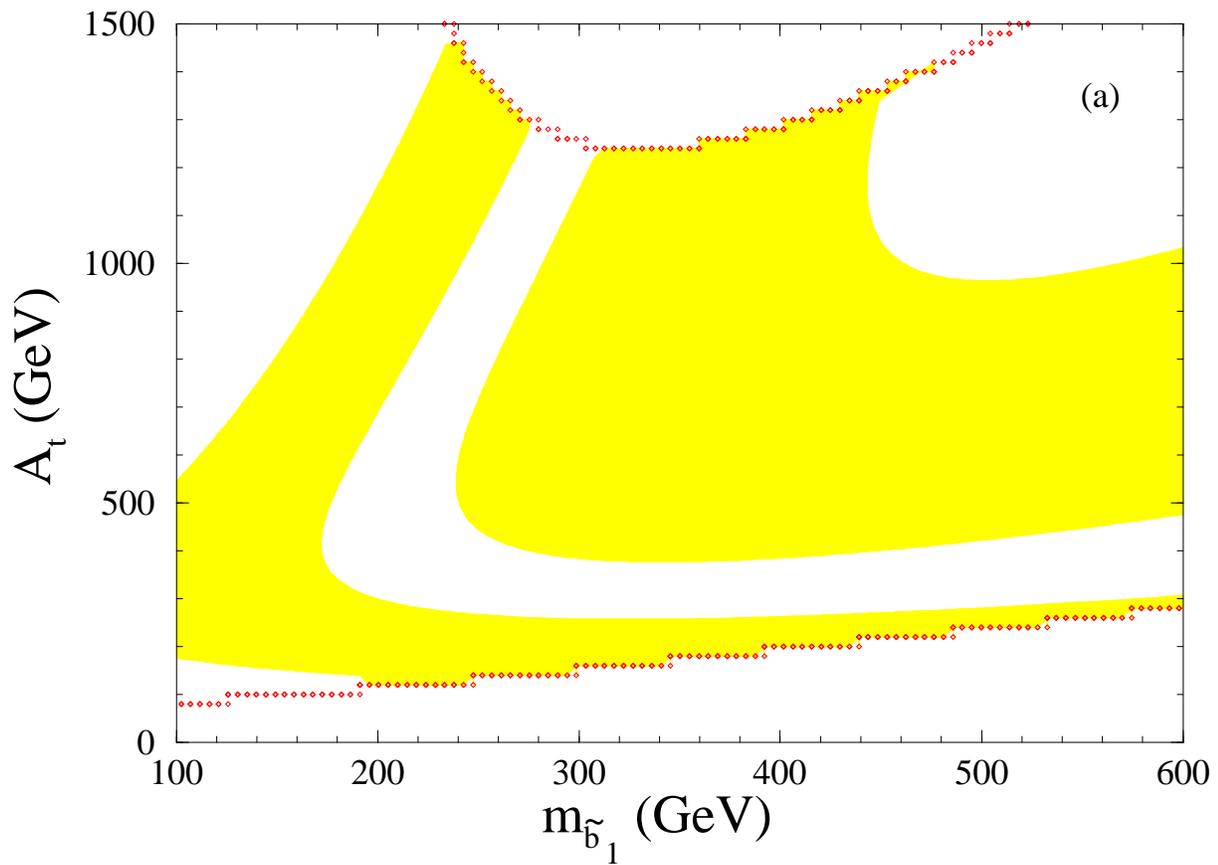
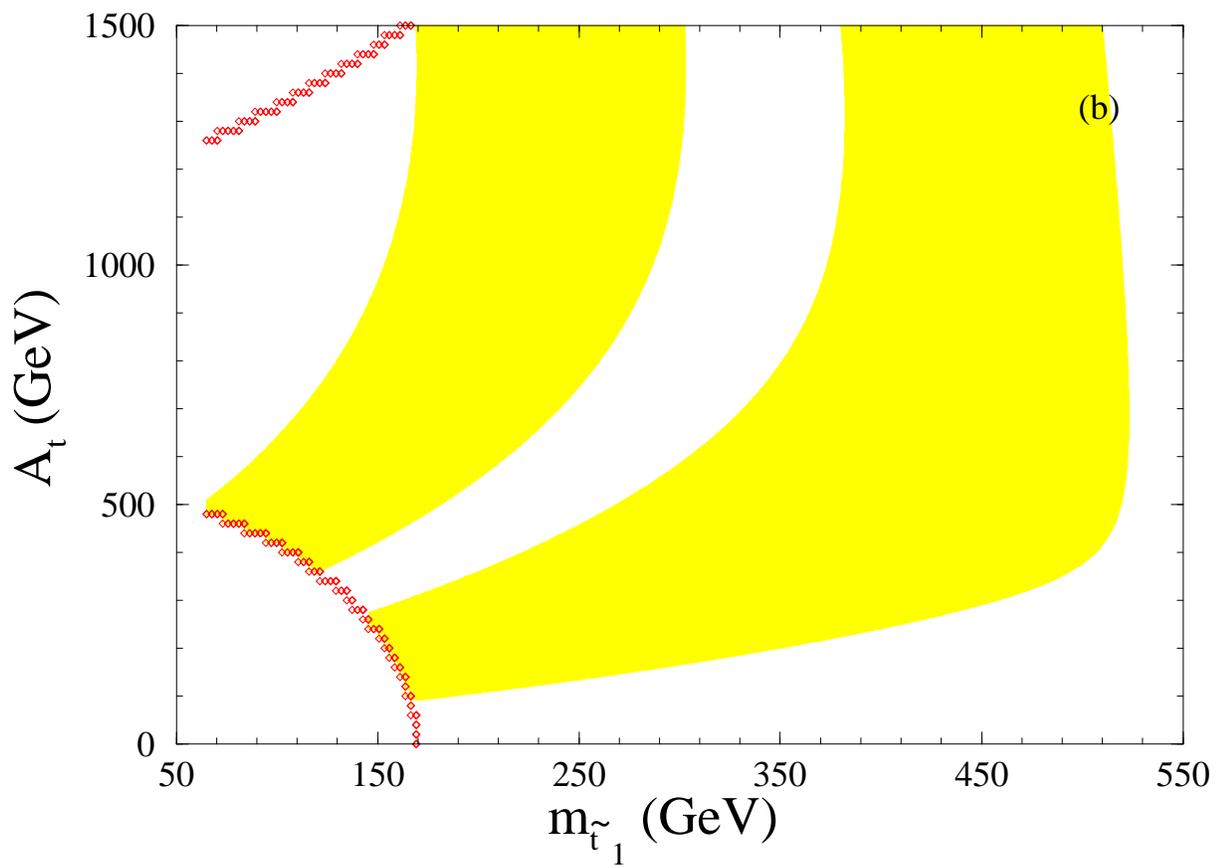
\item{\bf Fig.3} {\bf (a)} As in Fig.2, but 
in the $(m_{\tilde{b}_1},A_t)$-plane; {\bf (b)} As before, but in
$(m_{\tilde{t}_1},A_t)$-plane. Remaining inputs as in Fig.\,2.

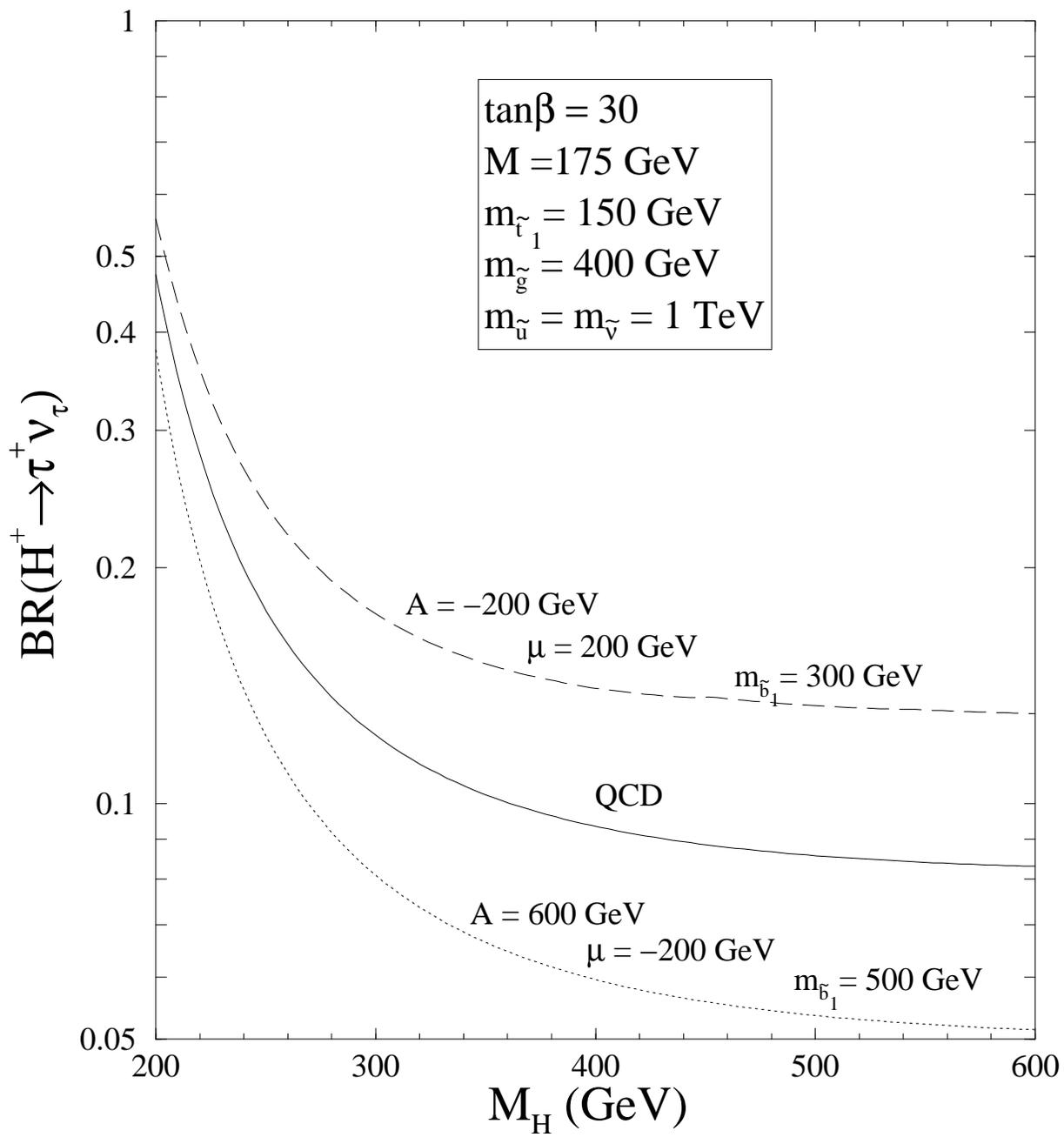
\item{\bf Fig.4}  The branching ratio of 
$H^+\rightarrow \tau^+\,\nu_{\tau}$ for positive
and negative values of $\mu$ and $A_t$ allowed by eq.(\ref{eq:bsg}),
as a function of the charged Higgs mass; $A$ is a common value for
the trilinear couplings.  The central curve includes
the standard QCD effects only.

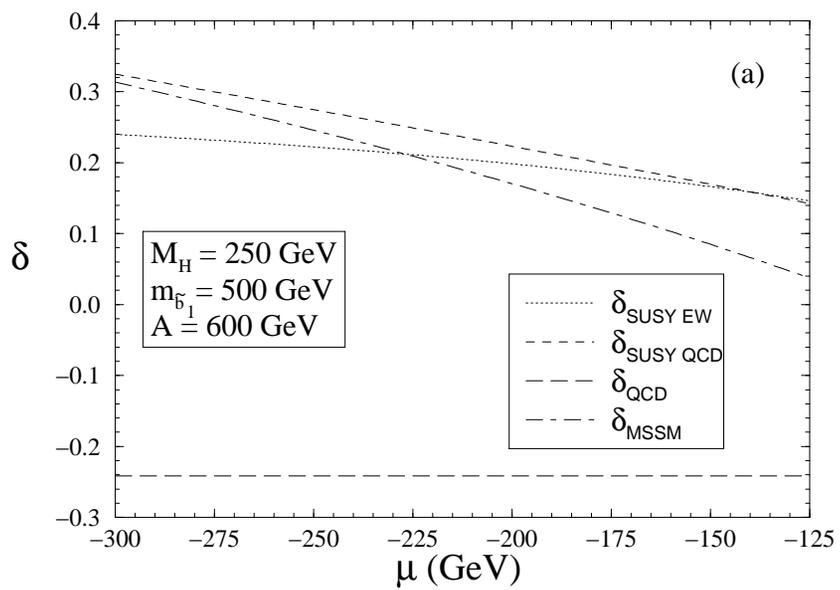
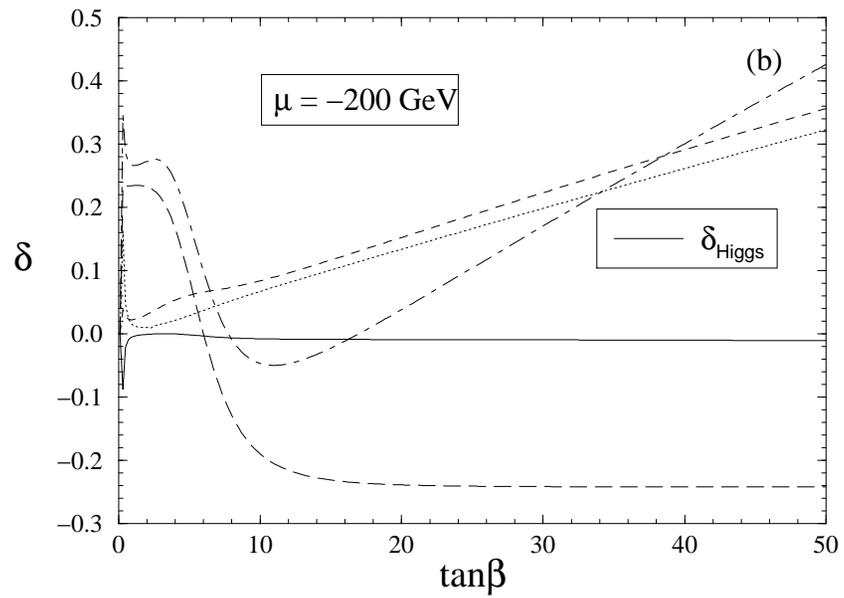
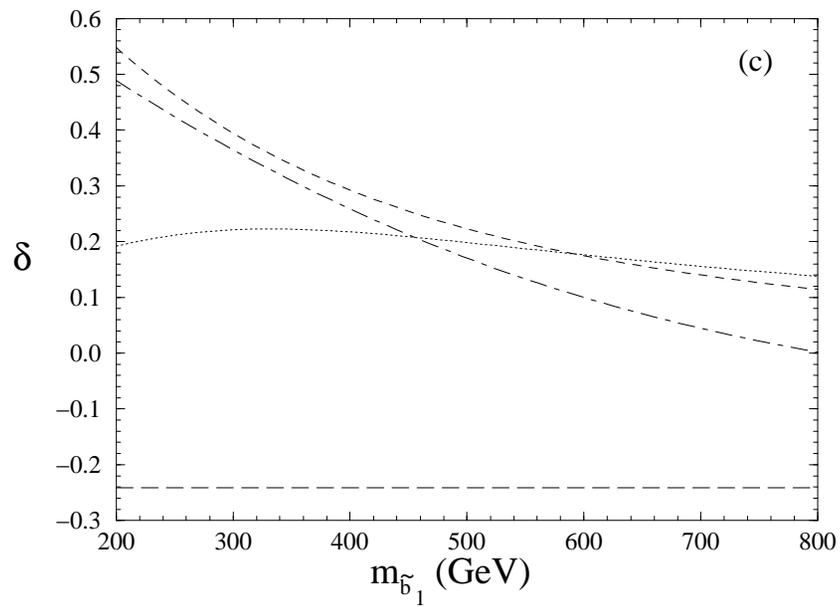
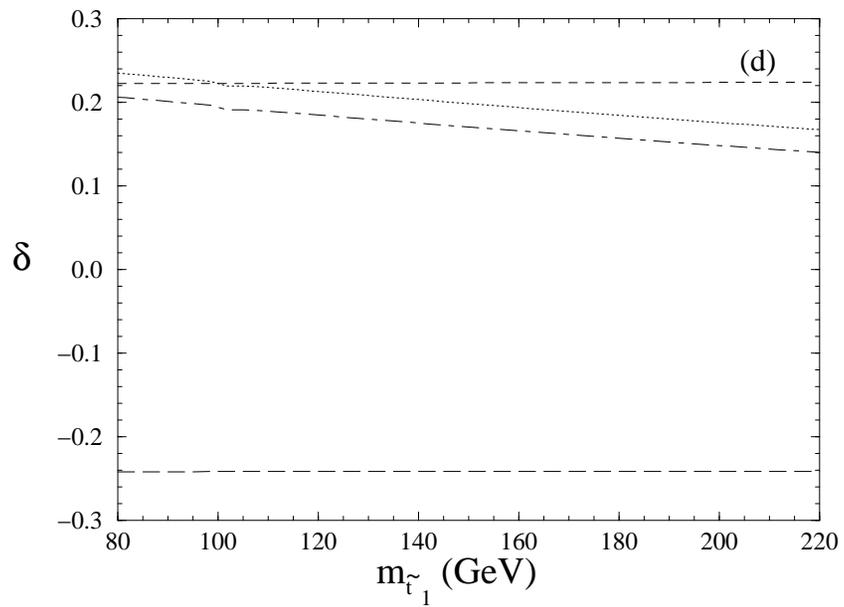
\item{\bf Fig.5} {\bf (a)} The SUSY-EW, SUSY-QCD, standard QCD
and full MSSM contributions to $\delta$, eq.(\ref{eq:deltag}), as a function of
$\mu$; {\bf (b)} As in (a), but as a function of
$\tan\beta$. Also shown in (b) is the Higgs contribution, $\delta_{Higgs}$;
{\bf (c)} As in (a), but as a function of
$m_{\tilde{b}_1}$; {\bf (d)} As a function of
$m_{\tilde{t}_1}$. Remaining inputs as in Fig.\,4.
 
\end{itemize}

\end{document}